\documentclass[a4paper,10pt]{article}
\usepackage{graphicx}

\begin{document}


\title{The role of AGN jets and intracluster magnetic fields in the formation and acceleration of cosmic rays\footnote{Published in Proceedings IAU Symposium No. 275, Romero, G.E., Sunyaev, R.A. and Belloni, T. (eds.)}}

\author{Gizani A.B. Nectaria$^1$\\
$^1$ Physics Laboratory, School of Science and Technology, Hellenic Open University, \\ Patra, Greece, email: {\tt ngizani@eap.gr} \\}

\date{}
\maketitle

\paragraph{Abstract}

 Using radio and X-ray data of two powerful radiogalaxies we attempt to
find out the role that radio jets (in terms of composition and power) as well as
intracluster magnetic fields play in the formation, propagation and acceleration of
cosmic rays. For this study we have selected the powerful radio galaxies Hercules A and 3C\,310 because of the presence of ring-like features in their kpc-scale radio emission instead of the usual hotspots. These two FR1.5  lie at the center of galaxy cooling flow clusters in a dense environment. 

We observed the unique jets of Hercules both in kpc- (multifrequency VLA data) and pc-scales (EVN observations at 18~cm). We have also observed the core and inner jets of 3C310 at 18~cm using global VLBI. 
We report on the work in progress. \\


\section{Introduction}

Powerful nearby AGN (central engine/shocks/jets) are correlated with ultra high energy cosmic rays (UHECRs) (\cite{dremmer}; \cite{ensslin96}). The intracluster medium (ICM) consists of relativistic particles (cosmic rays), weak magnetic fields and hot gas. In the current paper we report on the powerful radio galaxies Hercules A (z=0.154) and 3C\,310 (z=0.054) chosen because of the presence of ring-like features in their kpc-scale radio emission instead of hotspots. We adopt H$_{\circ}$ = 65 km s$^{-1}$ Mpc$^{-1}$ and q$_{\circ}$ = 0 throughout.

Hercules A and 3C\,310 present many similarities: They have double optical nuclei of similar absolute magnitude in R-band at low redshift. They are 
 hosts of galaxy clusters with similar gas temperature and a contribution from a point source. They are FR1.5 and have sharply bounded double lobes. Their lobes present asymmetry with respect to brightness, depolarisation and spectral index. They have no compact hotspots. Instead they contain radio ring-like features. Other high-brightness structure is also present with flatter 
spectra than the surrounding diffuse lobes. This may suggest a renewed outburst from the active nucleus. Their projected B-field follows closely the edges of the rings and lobes. They are both old sources with a steep spectral index$\alpha \simeq$ -1.4 (assuming that the flux density is given by S$_\nu \propto \nu^{\alpha}$). Their thermal pressure at the distance of the radio lobes is greater than the lobe minimum pressure. 

\section{Results and Discussion}

Radio spectral observations indicate the ageing of the emitting particles. The lobes of Hercules A have steeper spectrum than found in typical radio sources, and steepens further towards the centre (\cite{gizanir}). Its kpc-jets and rings have flatter spectrum than the surroundings strongly suggesting a recently renewed outburst from the active nucleus. The kpc-jets of Her A are well collimated at short distances from the core within the lobes, before dramatically disrupting. Both sources seem to have a large misalignment between pc-, and kpc-jets maybe suggesting a dense environment (\cite{gizanie}).   

Cosmic rays (CRs) could be trapped in the intracluster magnetic field in a similar manner to the interstellar magnetic field in the Galactic plane. Using the thermal energy content of the two clusters (\cite{gizanix}) and Faraday rotation measures for Hercules A (\cite{gizanir}) we have found a central value of 3 $\leq$ B$_\circ (\mu$G) $\leq$ 9 on tangling scale size 4$\leq$ l$_\circ$ (kpc) $\leq$ 35 while IC arguments suggest $\sim 4.3 \mu$G and $\sim 3.6 \mu$G for 3C\,310. The results are similar to the ones for clusters with host radio galaxy and a cooling flow, implying a nonthermal phase.  The other phase consists of cosmic ray protons, that have cooling times equal to or larger than the Hubble-time.  

The projected magnetic field closely follows the edges of the lobes, jets and rings of Her A. The result suggests that the lobe interior should be dominated by particle pressure, whereas the magnetic pressure should dominate in the shell region defined by the lobe boundary. In addition the gas thermal pressure is greater than the lobe minimum pressure implying confinement of the radio structure of the AGN by the ambient ICM rather than by shocks (\cite{gizanix}; \cite{leahy}). There is little entrainment, so the energy supply of the lobes mostly comes from relativistic particles and magnetic fields. Minimum pressure estimates imply a severe underestimate of the energy content of the jets of both sources. The magnetic field should be below equipartition and therefore unimportant in the lobe dynamics. Hence "invisible" particles (relativistic protons, low energy  e$^{-}$ / e$^{+}$) should dominate (\cite{gizanir};\cite{leahy}). 

For energy input from the host RG into the central region of clusters $\simeq 0.7 \cdot 10^{22} W kpc^{-3} h_{65}^{-2}$ (\cite{ensslin96}) the central cosmic ray energy $\epsilon_{CR(r)}$ estimated from the thermal energy density is (2.4-4.8)$\times 10^{-12}$ m$^{-3}$ Ws$^{-1}$ for Her A and 2.4 $\times 10^{-12}$ m$^{-3}$ Ws$^{-1}$ for 3C\,310. The injected jet power may dissipate and heat the gas, or could accumulate and support the ICM (magnetic fields and particles). The production rate for gamma rays above 100 MeV by $\pi_{\circ}$-decay after hadronic interactions of the energetic protons with the background gas is estimated to be (0.94 - 1.9)$\times 10^{-20}$ m$^{-3}$ s$^{-3}$ for Her A and 0.19$\times 10^{-20}$ m$^{-3}$ s$^{-3}$ for 3C\,310.

The short cooling time of the emitting CR electrons (suggested by the steepening of the spectral index which further implies short lifetimes of the radiating particles) and the large extent of the radio sources suggest an ongoing acceleration mechanism and also, to some extent, re-acceleration of the electrons and hence energy redistribution in the ICM.

\end{document}